\documentclass[12pt]{article}
\usepackage{amsmath}
\usepackage{amssymb}

\def\rdots{\mathinner{\mkern1mu\raise1pt\vbox{\kern1pt\hbox{.}}\mkern2mu
\raise4pt\hbox{.}\mkern2mu\raise7pt\hbox{.}\mkern1mu}}
\newcommand{\Z}{{\rm Z\kern-.35em Z}} \newcommand{\bP}{{\rm I\kern-.15em P}}
\newcommand{\Q}{\kern.3em\rule{.07em}{.65em}\kern-.3em{\rm Q}}
\newcommand{\R}{{\rm I\kern-.15em R}} \newcommand{\h}{{\rm I\kern-.15em H}}
\newcommand{\LC}{\kern.3em\rule{.07em}{.40em}\kern-.3em{\rm C}}
\newcommand{\C}{\kern.3em\rule{.07em}{.65em}\kern-.3em{\rm C}}
\newcommand{\T}{{\rm T\kern-.35em T}} \newcommand{\D}{{\kern-.5em /}}
\newcommand{\we}{\wedge} \newcommand{\ve}{\varepsilon} 
\newcommand{\ga}{\gamma} \newcommand{\om}{\omega} \newcommand{\pa}{\partial}
 \newcommand{\ra}{\rightarrow} 
\newcommand{\Om}{\Omega} \newcommand{\de}{\delta} 
 \newcommand{\al}{\alpha}
 \newcommand{\sig}{\sigma}
 \newcommand{\Ga}{\Gamma}

\begin{document}

\openup 1.5\jot

\centerline{\Large{The Interaction of Gravity with Other Fields}}

\vspace{.20in}
\centerline{by Joel A. Smoller$^*$}

\vspace{1in} \underline{Abstract}.  We consider the interaction of gravity, as
expressed by Einstein's Equations of General Relativity, to other force fields. 
We describe some recent results, discussing both the mathematics, and the
physical interpretations.  These results concern both elementary particles, as
well as cosmological models.  (This paper describes joint work variously done
with with F. Finster, N. Kamran, B. Temple, and S.-T. Yau.)

\vspace{.80in} \noindent \underline{Contents}:

\noindent 1.  Introduction

\noindent 2.  Background Material

\noindent 3.  Einstein-Dirac-Yang/Mills Equations

\noindent 4.  Decay Rates and Probability Estimates for Massive Dirac Particles
in a Charged, Rotating Black-Hole Background

\noindent 5.  Shock-Waves, Cosmology, and Black-Holes

\noindent \underline{ \ \ \ \ \ \ \ \ \ \ \ \ \ \ \ \ \ \ \ \ \ \ \ }

$^*$This work was partially supported by the NSF, Contract Number DMS-0103998

\section{Introduction}

\indent

We wish to describe how General Relativity modifies classical physics on two
different scales: the scale of elementary particles, and on cosmological scales,
the large scale structure of the Universe.  We shall discuss a few different
scenarios, describing the main ideas.  The more detailed mathematical
discussions can be found in our original papers.

The mathematics of General Relativity (GR) has several features which are
extremely interesting, and are quite different from the classical Newtonian
theory of gravity.  We will show how rich a subject GR is, both mathematically,
and in its physical applications and interpretations.  Indeed, GR is most
interesting from a physical viewpoint because it gives a different understanding
of physical phenomena.

The main feature of Einstein's Theory of GR is that the gravitational field is
the metric in 4-dimensional spacetime.  Thus GR is actually a theory of
spacetime in the sense that the underlying spacetime continuum is not fixed, but
is allowed to vary; indeed, massive bodies modify the curvature of their
surrounding spacetime geometry, and free particles traverse along geodesics
determined by the metric.  The second important feature of GR is the existence
of black holes; this notion has no classical analogue.  We shall see the role
played by black holes in various aspects of this paper.  Finally, we remark that
GR in and of itself, is a very beautiful subject.  This was already recognized
by Einstein himself who in 1915, in his presentation of the theory of GR to the
Prussian Academy in Berlin stated, ``Hardly anyone who has grasped the theory,
will be able to escape from its magic."  This is undoubtedly still true today.

\section{Background Material}

\ \ \ \ \ In this section we shall review some basic ideas in GR, Yang-Mills
(YM) equations and the Dirac (D) equation in four-dimensional Lorentzian
spacetime.

\vfill\eject

\noindent A. \underline{General Relativity}

General Relativity is Einstein's Theory of Gravity, and is based on the
following 3 hypotheses.

\begin{itemize}
\item [(E$_1$)] The gravitational field is the metric $g_{ij}$ in 4-dimensional
spacetime.  The metric is assumed to be symmetric: $g_{ij} = g_{ji}, \ \ i,j = 0
- 3$.  \item [(E$_2$)] At each point, $g_{ij}$ can be diagonalized as $$g_{ij} =
\ {\rm diag}(-1,1, 1, 1).$$
\item [(E$_3$)] The equations should be independent of the particular
coordinates involved, and thus the equations should be tensor equations.
\end{itemize}

$(E_1)$ is Einstein's brilliant insight, whereby he ``geometrizes" the
gravitational field, replacing one Newtonian potential for the gravitational
field by the ten metric potential functions, $g_{ij}, \ i \le j$.  $(E_2)$ means
that Special Relativity is included in GR, (and also that the metric tensor is
everywhere non-singular),while $(E_3)$ asserts that coordinates are merely an
artifact, and physics shouldn't depend on the choice of coordinates.

The metric $g_{ij} = g_{ij}(x), \ \ i,j = 0 - 3, \ \ x = (x^0, x^1, x^2, x^3), \
\ x^0$ = ct (c = speed of light), is a tensor defined on 4-d spacetime.  {\it
Einstein's equations} are ten (tensor) equations for the unknown metric $g_{ij}$
(gravitational field), and take the form
\begin{equation} R_{ij} - \frac 1 2 \ Rg_{ij} = \sig
T_{ij} \ .  \end{equation}
The left-hand side $G_{ij} = R_{ij} - \frac 1 2 \ Rg_{ij}$ is
the {\it Einstein tensor} and is a {\it geometric} quantity, depending only on
$g_{i j}$ and its derivatives, while $T_{ij}$, the {\it energy-momentum tensor},
represents the source of the gravitational field, and encodes the distribution
of matter.  The word ``matter" in GR refers to everything which can produce a
gravitational field, including elementary particles and electromagnetic fields. 
Since the Einstein tensor {\it identically} satisfies the equation $G^i_{j;i} =
0$ (the covariant divergence vanishes) it follows that on solutions of (1),
$T^i_{j;i} = 0$, and this in turn expresses the laws of conservation and
momentum, [23].  The quantities which make up the Einstein tensor $G_{ij}$ are
given as follows.  First from the metric tensor $g_{i j}$, we form the {\it
Levi-Civita connection} ({\it Christofffel symbols}) $\Ga^k_{ij}$:
\[ \Ga^k_{ij} = \frac 1 2 \ g^{k\ell} \left( \frac{\pa g_{\ell j}}{\pa x^i} +
\frac{\pa g_{i \ell }}{\pa x^j} - \frac{\pa g_{i j}}{\pa x^\ell} \right), \ \ \
\ \ i,j,k = 0 - 3, \]
where $[g^{k\ell}] = [g_{k\ell}]^{-1}$, and summation convention is
employed; namely every up-down index is to be summed from 0 - 3.\footnote{In a
letter to a friend Einstein wrote, ``I have made a great discovery in
mathematics; I have suppressed the summation sign every time the summation must
be made over an index that occurs twice ...".  } Using these $\Ga^k_{ij}$, we
construct the {\it Riemann curvature tensor} $R^i_{qk\ell}$:
\[ R^i_{qk\ell} = \frac {\pa \Ga^i_{q\ell}}{\pa x^k} - \frac {\pa \Ga^i_{qk}}{\pa
x^\ell} + \Ga^i_{pk} \Ga^p_{q\ell} - \Ga^i_{p\ell} \Ga^p_{qk}.  \]
Finally, we can explain the terms $R_{ij}$ and $R$ in $G_{ij}$; namely $R_{ij} =
R^s_{isj}$ is the {\it Ricci tensor} and $R = g^{ij} R_{ij}$ is the {\it scalar
curvature}.

The quantity $\sig$ is a universal constant defined by \[ \sig =
\frac{8\pi\kappa}{c^4} \] where $\kappa$ is Newton's gravitational constant, and
$c$ is the speed of light (both in ``appropriate" units; we shall often choose
units in which $\kappa = 1 = c$).

From these definitions, one immediately sees the enormous complexity of the
Einstein equations (2.1).  For this reason, we shall often seek solutions which
have a high degree of symmetry, the aim being to make the resulting equations
mathematically tractable.

\noindent B. \underline{Black Hole Solutions}

Consider the gravitational field outside of a ball of mass $M$ in $\R^3$. 
Solving Einstein's equations $G_{ij} = 0$, gives the celebrated {\it
Schwarzschild solution} (1916): \vfill\eject
\begin{equation} ds^2 = - \left( 1 - \frac{2m} r
\right) c^2 dt^2 + \left( 1 - \frac{2m} r \right)^{-1} dr^2 + r^2 d\Om^2,
\end{equation}
where $m = \frac{\kappa M}{c^2}$, and $d\Om^2 = d\theta^2 + \sin^2 \theta
d\phi^2$ is the standard metric on the unit 2-sphere.  Since $2m$ has dimensions
of length, it is called the {\it Schwarzschild radius}.  Observe that on the
sphere $r=2m$, the metric becomes singular; indeed $g_{tt} = 0$ and $g_{rr} =
\infty$.  By transforming the metric (2.2) to so-called {\it Kruskal
coordinates}, [1], one sees that the Schwarzchild sphere $r=2m$, has the
physical characteristics of a {\it black hole}: light and nearby particles can
enter, nothing can escape, and there is an intrinsic (non-removable) singularity
at the center $r=0$.

More generally, we could consider a metric of the form
\begin{equation} ds^2 = - T(r)^2 dt^2
+ A^{-1}(r) dr^2 + r^2d \Om^2, \end{equation}
where \[ A(r) = 1 - \frac{2m(r)} r, \ \ \ \
2m(r) = r(1-A(r)); \] $m(r)$ is called the ``mass" function.  For this metric,
we define a {\it black hole} solution to be a solution of Einstein's equations
which satisfies \[ A(\rho) = 0, \ \ {\rm for \ some} \ \ \rho > 0, \] and $A(r)
> 0$ if $r > \rho$.  $\rho$ is the radius of the black hole, and is often
referred to as the {\it event horizon}.  (Defined by the condition that it be
the largest zero of the $dr^2$ term in the metric.)

\noindent C. \underline{Yang-Mills Equations}

The Yang-Mills equations are a generalization of Maxwell's equations of
electromagnetism.  To see this, we first write Maxwell's equations in an
invariant way.  For this, let $A$ be the scalar-valued 1-form \[ A = A_i dx^i \
\ , \ \ \ \ \ A_i \in \R \ , \] called the {\it electromagnetic potential} (by
physicists), or a {\it connection} (by geometers).  Then, $F$, the
elecromagnetic field (curvature) is the 2-form defined by
\begin{equation} F = dA .  \end{equation}
In coordinates, $F$ can be written as \[ F = F_{ij} dx^i \we dx^j \ , \ \ \ \ \
F_{ij} = \frac{\pa A_i}{\pa x^j} - \frac{\pa A_j}{\pa x^i}.  \] In this set-up,
Maxwell's equations take the form \[ dF = 0 \ , \ \ \ \ \ d^*F = 0.  \] The
first equation follows trivially from (2.4), while for the second equation, the
``star" is called the Hodge star operator and in four dimensions, it maps
2-forms to 2-forms and is defined by \[ (^* F)_{k\ell} = \frac 1 2 \ \sqrt{|g|}
\ve_{ijk\ell} F^{ij}, \] where $g=\det(g_{ij})$, and $\ve_{ijk\ell}$ is the
completely anti-symmetric symbol defined as $\ve_{ijk\ell} = \ {\rm
sgn}(i,j,k,\ell)$.  Here, as usual, we always use the metric to raise (or lower)
indices, so that \[ F^{ij} = g^{\ell i} g^{mj} F_{\ell m} .  \] It is important
to notice that $^*F$ depends on metric $g_{ij}$.

The {\it Yang-Mills} equations are a set of equations which generalize Maxwell's
equations.  Thus, to each Yang-Mills equation is associated a Lie group $G$ (of
symmetries), called the {\it gauge group}.  For such $G$, we consider its Lie
algebra $\frak{g}$, defined as being the tangent space at the identity of $G$. 
Now suppose that $A$ is a $\frak g$-valued 1-form; i.e. \[ A = A_i dx^i, \]
where each $A_i $ is in $ \frak{g}$.  In this (slightly!)  more general case,
the curvature 2-form $F$ is defined by \[ F = dA + A \wedge A, \] and in
components, \[ F_{ij} = \frac {\pa A_j}{\pa x^i} - \frac {\pa A_i}{\pa x^j} +
[A_i, A_j], \ \ \ \ A_i \in \frak{g}.  \] Notice that the commutator, $[A_i,
A_j] = 0$ if $G$ is an abelian group, but is generally non-zero if $G$ is a
matrix group.  The Yang-Mills equations are again given by
\begin{equation} dF = 0 \ \ \ \ \
{\rm and} \ \ \ \ \ d^*F = 0.  \end{equation}
These generalize Maxwell's equation because
for Maxwell's equation, $G = U(1)$, the circle group $(U(1) = \{ e^{i\theta} :
\theta \in \R \})$, so $\frak{g}$ consists of scalars and the commutator term
vanishes.  Note that for Maxwell's equations $d^*F = 0$ is a {\it linear}
equation for the $F_{ij}$'s, or for the unknown ``connection coefficients"
$A_i$.  However if $G$ is non-abelian, say $G=SU(2)$, then the Yang-Mills
equations $d^*F=0$ become {\it non-linear} equations for the (unknown) matrices
$A_i$.
 
\noindent D. \underline{The Dirac Equation in Curved Spacetime}

The Dirac equation brings in {\it Quantum Mechanics}, and is a generalization of
Schr\"{o}dinger equation to the relativistic case.  It also describes the
intrinsic ``spin" of certain elementary particles.  The Dirac equation takes the
form
\begin{equation} (G - m)\Psi = 0, \end{equation}
where $G$ is the Dirac operator and $\Psi$ is a
complex-valued 4-vector called wave the function (spinor) of a
fermion\footnote{Fermions are distinguished from bosons in that fermions have
half-integral multiples of spin, while bosons have integral multiples of spin. 
Only fermions obey the Pauli Exclusion Principle.  Examples of fermions are
protons and antiprotons, neutrinos, electrons and positrons, quarks, and
leptons.  Examples of bosons are mesons, photons, and pions.} (Dirac particle)
of mass $m$.  The Dirac operator $G$ can be written as \[ G = i G^j(x) \frac
\pa{\pa x^j} + B(x) \] where the $G^j$ are $4 \times 4$ matrices called the {\it
Dirac matrices} and $B$ is a $4\times 4$ matrix.  The Dirac matrices $G^j$ and
the Lorentzian metric $g_{ij}$ (defined on 4-d spacetime) are related by
\begin{equation}
g^{jk} I = \frac 1 2 \Big\{ G^j, G^k \Big\}, \end{equation}
where $\Big\{ G^j, G^k \Big\}$
is the {\it anti-commutator} \[ \Big\{ G^j, G^k \Big\} = G^jG^k + G^kG^j .  \]
The Dirac matrices thus depend on the metric.  

Now let $H$ be any space-like hypersurface in $\R^4$, with future directed
normal vector field $\nu = (\nu_i)$, and let $d\mu$ be the invariant measure on
$H$ induced by the metric $g_{ij}$.  Define a scalar product on solutions,
$\Psi, \Phi$ of the Dirac equation by \[ \Big< \Psi|\Phi \Big> = \int_H \
\bar\Psi G^j\Phi \nu_j d\mu \ .  \] This scalar product is positive definite,
and, as a consequence of {\it current conservation} (c.f. [4]) \[ \nabla_j
\bar\Psi G^j\Phi = 0, \] it is independent of the choice of the hypersurface
$H$.  By direct generalization of the expression \[ \bar\Psi \ga^0 \bar\Psi =
|\Psi|^2 \] in flat Minkowski space given originally by Dirac, ([2]), where \[
\ga^0 = \left( \begin{array}{cc} {\bf 1} & 0 \\
 \\
0 & -{\bf1} \end{array} \right), \] in our case $\bar\Psi$ is the adjoint
spinor, defined by \[ \bar\Psi = \Psi^* \left( \begin{array}{cc} {\bf 1} & 0 \\
 \\
0 & -{\bf 1} \end{array} \right), \] ${\bf 1}$ is the $2\times2$ identity
matrix, and $\Psi^*$ denotes complex conjugation,
\begin{equation} \bar \Psi G^j \Psi \nu_j
\end{equation}
is interpreted to be the {\it probability density} of the Dirac particle. 
We normalize solutions of the Dirac equation by requiring
\begin{equation} \Big< \Psi |
\bar\Psi \Big> = 1.  \end{equation}

If the Lorentzian metric admits a black-hole solution at $r = \rho > 0$, (2.9)
is replaced by
\begin{equation} \begin{array}[t]{c} {\displaystyle\int} \\
 {\scriptstyle \{t={\rm const.}, \; r > r_0\}} \end{array} \bar\Psi G^j \Psi
 \nu_j d\mu < \infty \end{equation}
 for all $r_0 > \rho$.  For normalized solutions of the
 Dirac equation, the integral (2.10) gives the probability for the Dirac
 particle to be in the region $r > r_0$.  We shall only consider Dirac particles
 which lie outside the event horizon.  This is because the probability density
 (2.8) is not necessarily positive inside the event horizon so that a positive
 infinite contribution of the integral outside the event horizon can be
 compensated by a negative infinite contribution inside the event horizon.  We
 thus demand that the integral (2.10) away from, and outside of the event
 horizon, be finite.

In the paper [4], it is shown that the Dirac matrices $G^j$ can be chosen to be
any $4\times 4$ matrices which are Hermitian with respect to the scalar product
\[ \Big< \Psi|\Phi \Big> = \int_{\R^4} \bar\Phi \Psi \sqrt{|g|} \; dx \] and
satisfy the anti-commutation relations (2.7).  This gives us more flexibility in
choosing the Dirac matrices.  Namely, the relations (2.7) do not uniquely
determine the Dirac matrices in curved spacetime.  However it was proved in [4]
that all choices of Dirac matrices satisfying (2.7) yield unitarily equivalent
Dirac operators.

With this introduction to the Einstein equations, the YM equations and Dirac's
equation, we turn to our first topic, the coupled Einstein-Dirac-Yang/Mills
(EDYM) equations.

\bigskip \bigskip

\section{Einstein-Dirac-Yang/Mills Equations}
\centerline{(joint work with F. Finster and S.T.-Yau)}
\setcounter{equation}{0}

\ \ \ \ The EDYM equations are obtained by varying the action \[ S = \int \left[
\frac 1{16\pi \kappa} \; R + \bar\Psi(G-m)\Psi - \frac 1{16\pi e^2} \ {\rm
Tr}(F_{ij}F^{ij})\right] \sqrt{|g|} \ dx \] over Lorentzian metrics $g_{ij}$,
$\frak{g}$-valued Yang/Mills connections $A_i dx^i$ (where $\frak{g}$ is the Lie
algebra of the given gauge group), and 4-spinors $\Psi$.  Here $e$ is the
so-called Yang/Mills coupling constant, and $\kappa$ is Newton's gravitational
constant.  In general, the EDYM equations are extremely complicated.  We
specialize here to {\it spherically symmetric solutions} depending only on the
radius $r = |x|$.  Taking the gauge group as $SU(2)$\footnote{The gauge group
$SU(2)$ corresponds to the weak nuclear force, while $SU(3)$ corresponds to the
strong nuclear force.} together with a special ansatz for the spinors, we can
simplify the Dirac equations from complex 4-spinors to the case of an equation
for real 2-spinors $(\al, \beta)^t$, [8].  In this set-up the (SU(2)) EDYM
equations take the following form:
\begin{eqnarray}
\sqrt{A} \left( \begin{array}{c} \alpha \\
 \\
 \beta \end{array} \right)' &=& \left(
\begin{array}{cc}
W/r & - (m+\om)T \\
 \\
\bigskip -m+\om T & - W/r \end{array} \right) \left( \begin{array}{c} \alpha \\
 \\
\beta \end{array} \right) \\
\nonumber \\
rA' &=& 1 - A - \frac {\kappa}{e^2} \frac{(1-W^2)^2}{r^2} -2\kappa\om
T^2(\alpha^2 + \beta^2) - \frac {2\kappa} {e^2} A (W')^2 \\
 \nonumber \\
2rA \frac{T'}T &=& -1+A+\frac {\kappa}{e^2} \frac{(1-W^2)^2}{r^2} + 2\kappa
mT(\alpha^2 - \beta^2) -2\kappa \om T^2(\alpha^2 + \beta^2) \\
&+& 4\kappa \frac T r \; W\al \beta - \frac{2\kappa}{e^2} A (W')^2 \nonumber \\
 \nonumber \\
 r^2 AW'' &=& -(1-W^2)W + e^2 r T\al \beta - r^2 \ \frac{A'T-2AT'}{2T}\ W'.
\end{eqnarray}

\bigskip \noindent Here $m$ is the rest mass of the Dirac particle, $\om$ is its
energy, and $W$ is the unknown connection coefficient.  Equation (3.1) is the
Dirac equation, (3.2) and (3.3) are the Einstein equations (for the Lorentzian
metric of the form (2.3)), while (3.4) is the Yang/Mills equation.  Notice that
equations (3.1), (3.3) and (3.5) all become singular when $A=0$; i.e., at a
black-hole.  (This is special to Einstein's equations, and is the first
significant difference that we see from classical equations.)

Smooth solutions of (3.1) - (3.4) are those which are defined for all $r \ge 0$,
and are called {\it particle-like solutions}.  Black hole (BH) solutions
correspond to solutions having 
$A(\rho) = 0$ for some $\rho > 0$, and are defined for all $r > \rho$,
with $A(r) > 0$, if $r >\rho$. 
In addition solutions of (3.1)-(3.4) are required to satisfy certain initial
condictions, (see [8, 9]), together with the following global conditions :
\begin{equation}
\int^\infty_0 (\al^2 + \beta^2) \frac T{\sqrt{A}} \ dr = 1 \ \ \ \ \ \ ({\rm for
\ particle-like \ solutions)}, \end{equation}
\begin{equation} 0 < \int^\infty_{r_0} (\al^2 + \beta^2)
\frac T{\sqrt{A}} \ dr < \infty \ \ \ \ \ \ {\rm for \ each} \ \ r_0 > \rho, \end{equation}
(for BH solutions),
\begin{equation} \lim_{r \ra \infty} r(1-A(r)) < \infty \ , \end{equation}
\begin{equation}
\lim_{r \ra \infty} \; T(r) = 1 \end{equation}
\begin{equation} \lim_{r \ra \infty} \Big( W(r),
W'(r))\Big), \ \ \ \ \ {\rm is \ finite}.  \end{equation}
Conditions (3.5) and (3.6) state
that the spinors (wave functions) are normalizable, (3.7) means that the
fermions have finite (ADM) mass, (3.8) (together with (3.7)) means that the
gravitational field is asymptotically Minkowskian, while (3.9) asserts that the
YM field is well-behaved.  We now state our main results, and discuss them
afterward.

\bigskip \noindent \underline{Theorem 3.1} ([8]): {\it There exist stable
particle-like solutions of the EDYM equations for arbitrarily weak gravitational
coupling constant:} $m^2k/e^2 \ra 0$.

\bigskip \noindent \underline{Theorem 3.2} ([9]): {\it Every black hole solution
of the EDYM equations cannot be normalized; namely, the spinors must vanish
identically outside of the black hole.}

\bigskip \noindent Both of these results are unexpected and rather surprising.

Theorem 3.1 is obtained using numerical methods.  This (stability) result has
been shown to be false for the (SU(2)), Einstein-Yang/Mills (EYM) equations,
[21, 22], and shows that Dirac particles in a gravitational field form bound
states if an additional strong coupling to a non-abelian Yang-Mills field is
taken into account.  In [8], we employ some new numerical techniques, including
a multi-parameter ``shooting method", and a scaling technique introduced in [5]. 
Our result shows that weak as gravity is (e.g., it is $10^{21}$ times weaker
than electromagnetism), it still has a regularizing effect.  Although gravity is
not renormalizable (which means that the problem cannot be treated in a
perturbation expansion), our solutions of the EDYM equations are regular and
well-behaved for arbitrarily weak gravitational coupling.  Our stability result
is obtained by using Conley Index techniques, [17].

The proof of Theorem 3.2 is given in [9].  The result shows that the EDYM
equations do not exhibit normalizable BH solutions.  Thus, in the presence of
quantum mechanical Dirac particles, static, spherically symmetric BH's do not
exist.  Another interpretation of our result is that it indicates that Dirac
particles can only either disappear into the BH or escape to infinity.  We shall
discuss this further in the next section.

\section{Decay Rates and Probability Estimates for \\ Massive Dirac Particles in
a Charged, \\ Rotating Black Hole Geometry}
\centerline{(joint work with F. Finster, N. Kamran, and S.-T. Yau)}
\setcounter{equation}{0}

\ \ \ \ \ Consider the physical situation of a Dirac particle (fermion,
electron, proton, etc.)  of mass $m>0$ {\it outside} of a charged, rotating
black hole.  Question: What is its long time behavior, as $t \ra \infty$?

	In the classical case (no Dirac equation, no quantum mechanical
	considerations), the particle can remain in a time-periodic orbit around
	the black hole (see [3]).  In this section we shall show that our
	previous ``indication", (mentioned at the end of the last section) is
	valid.  That is, if quantum mechanical effects are taken into account,
	the picture completely changes, and the Dirac particle either enters the
	black or hole, or tends to infinity; no other possibilities can occur.

	In the paper [7], we proved that there are no normalizable, time
	periodic solutions of the Dirac equation in a
	Reissner-Nordstr\"{o}m\footnote{The Reissner-Nordstr\"{o}m solution is
	an extension of the Schwazschild solution, and is a static solution of
	the coupled Einstein-Maxwell equations; see [1].} black-hole background;
	in particular there are no static solutions of the Dirac equation in
	such a background metric.  This result was extended in [10] to the case
	in which the background geometry is that of a charged, rotating black
	hole.  In this section we shall discuss such black holes and we shall
	show that our above ``indications" are correct; i.e., in such a
	background geometry, a Dirac particle either tends to infinity, or
	enters the black hole -- there are no other possibilities.  In addition
	we shall obtain probability estimates on which of these two
	possibilities occurs.  We begin with a brief discussion of the geometry
	of a charged rotating black hole.

In 1963, R. Kerr found a solution of Einstein's equations corresponding to a
rotating black hole [15].  This result was generalized by Kerr and Newman; (see
[11]), to a {\it charged} rotating black hole (called the Newman-Penrose
solution of Einstein's equations):
\begin{equation} ds^2 = \frac \Delta U (dt - a \sin^2 \phi
d \phi)^2 - U\left( \frac{dr^2}\Delta + d\theta^2 \right) - \frac{\sin^2\theta}U
\Big[ adt - (r^2+a^2)d\phi \Big]^2, \end{equation}
where
\begin{eqnarray*}
U(r, \theta) &=& r^2 + a^2 \cos^2 \theta \\
\Delta(r) &=& r^2 - 2Mr + a^2 + Q^2,
\end{eqnarray*}
and $M, \; aM$, and $Q$ denote the mass, angular momentum, and charge of the
black hole, respectively; $a$ is the angular velocity of the black hole.  From
our earlier remarks, the largest root of $\Delta$ corresponds to the event
horizon of the black hole.  We assume\footnote{This is the so-called
``non-extreme" black hole.}
\begin{equation} M^2 > a^2 + Q^2, \end{equation}
then \[ r = \rho \equiv M +
\sqrt{M^2 - a^2 - Q^2} \] defines the event horizon, the boundary of the
charged, rotating, black hole (CRBH).

Now consider the Cauchy Problem for a Dirac particle of mass $m > 0$, and charge
$q$, in this CRBH background geometry, (see [11], where we write the Dirac
equation in this different way, taking the spin connection into account) having
initial data outside of the event horizon $r = \rho$:
\begin{equation} (i\ga^j
D_j-m)\Psi(t,x) = 0 \end{equation}
\begin{equation} \Psi(0,x) = \Psi_0(x), \ \ \ \ \ |x| > \rho.  \end{equation}
Here is our first result.

\bigskip \noindent \underline{Theorem 4.1}: {\it Let $\de > 0$ be given, and let
$R > \rho + \de$.  Consider the (annular) region in $\R^3$ \[ K_{\delta,R} = \{
\rho + \de \le r \le R \}.  \] Then the probability for the Dirac particle to be
inside $K_{\de,R}$ tends to zero as $t \ra \infty$; namely,}
\begin{equation} \lim_{t \ra
\infty} \int_{K_{\de,R}} (\bar \Psi \ga^j \Psi)(t,x) \nu_j d\mu = 0.  \end{equation}

Thus with probability one, the Dirac particle leaves every bounded set in
$\R^3$.  The proof of this theorem is based on a result of Chandrasekhar, [3],
who showed that the Dirac equation in the CRBH geometry can be separated into
ODE's.  Moreover, using this, we can write the Dirac propagator as (see [11,
12])
\begin{equation} \Psi(t,x) = \frac 1\pi \sum_{k,n\in \Z} \int^\infty_{-\infty} e^{-i\om
t} \sum^2_{a,b=1} t^{k\om n}_{ab} \Psi^{k\om n}_a(x) \left< \Psi^{k\om n}_b |
\Psi_0 \right> d\om, \end{equation}
where the (positive) scalar product is defined in [11]. 
Here $k$ and $n$ are generalized angular momentum numbers (arising from
Chandrasekhar's separation theorem), $\Psi^{k\om n}_a$ are solutions of the
Dirac equation, and the $t^{k\om n}_{ab}$ are generalized transmission
coefficients.  These functions can be written in terms of the fundamental
solutions of the resulting ODE's.  For $|\om| > m$, near the event horizon, the
$\Psi^{k\om n}_a$ go over to spherical waves.  $\Psi^{k\om n}_1$ corresponds to
incoming waves (moving towards the black hole), and $\Psi^{k\om n}_2$
corresponds to outgoing waves (moving away from the black hole).  Near infinity,
the $\Psi^{k\om n}_a$ again go over to spherical waves.  For $|\om| < m$, the
fundamental solutions (for $a=1,2$) near the event horizon are linear
combinations of both incoming and outgoing waves.  $\Psi^{k\om n}_1$ (resp. 
$\Psi^{k\om n}_2$) decays (resp.  grows) exponentially at infinity.  The theorem
is proved using the representation (4.6).

Theorem 4.1 implies that the Dirac particle must leave every bounded set in
$\R^3$.  Equivalently the Dirac wave function $\Psi$ decays to zero in
$L^\infty_{loc}$, outside and away from the event horizon.  It follows that the
Dirac particle {\it must} eventually either disappear into the black hole, or
escape to infinity; these are the only possibilities.  This raises the following
questions:
\begin{itemize}
\item[(A)] What is the likelihood of each of these two possibilities?,
\end{itemize}
 and
\begin{itemize}
\item[(B)] What is the rate of decay of the Dirac wave function on a compact
subset of $\R^3$ outside of the black hole?
\end{itemize}

We consider question (B) first.  To this end, let $q$ denote the charge of the
Dirac particle and assume that the Dirac particle has ``small charge", in the
sense that\footnote{This means that the gravitational attraction is the dominant
force far from the black hole.  Indeed, at large distances from the black hole
the metric is asymptotically Minkowskian, and the gravitational and
electromagnetic fields can be described by the Newtonian limit.  In this limit
the gravitational and electrical forces are \ $ mM/r^2$ \ and \ $qQ/r^2$, \ the
Newton and Coulomb laws, respectively.}
\begin{equation} mM > |qQ|.  \end{equation}
We also assume that
the angular momentum of the initial data is {\it bounded}.  This means that the
summation in (4.6) is over a bounded set of integers $n$ and $k$; say \[ |k| \le
k_0 \ \ \ {\rm and} \ \ \ |n| \le n_0.  \] We then have the following theorem;
(see [12] for the proof).

\bigskip \noindent \underline{Theorem 4.2}: (Decay Rates): {\it Consider the
Cauchy problem (4.3), (4.4) for the Dirac equation in the non-extreme CRBH
background geometry, with small charge.  Assume that the Cauchy data $\Psi_0$ is
smooth with compact support outside of the event horizon, and has bounded
angular momentum.

\noindent {\rm (i)} \ \ If for any $k$ and $n$
\begin{equation} \limsup_{\om \searrow m}
\left| \left< \Psi^{k\om n}_2 | \Psi_0 \right> \right| \not= 0, \ \ {\rm or} \ \
\liminf_{\om \nearrow -m} \left| \left< \Psi^{k\om n}_2 \; | \; \Psi_0 \right>
\right| \not= 0, \end{equation}
then
\begin{equation} |\Psi(t,x)| = ct^{-5/6} + O(t^{-5/6 \; - \; \ve})
\ \ \ {\rm as} \ \ \ t \ra \infty, \end{equation}
where $c=c(x) \not= 0$ and any $\ve < \frac 1{30}$.

\noindent {\rm (ii)} \ \ If for all $k,n$, and $a=1,2, \ \ \left< \Psi^{k\om
n}_a \; | \; \Psi_0 \right> = 0$ for all $\om$ in a neighborhood of $\pm \; m$,
then for any fixed $x$, $\Psi(x,t)$ decays rapidly in $t$, (faster than any
polynomial).  }

For a quantum mechanical particle, the Heisenberg Uncertainty Principle implies
that its kinetic energy is not precisely known.  Since the associated
Hamiltonian has continuous spectrum (the operator is defined on an unbounded
interval), the energy $\om$ of the Dirac particle is a continuous parameter. 
Condition (4.8) means that the initial energy distribution has outgoing
components near $\om = m$ or near $\om = -m$.  Our theorem implies that in this
case, the decay rate is $t^{-5/6}$.  That is, the decay rate $t^{-5/6}$
quantifies the effect of the black hole's attraction, on the long time behavior
of the Dirac particle.

In flat Minkowski space solutions of the wave equation having compactly
supported data, decay rapidly (Hughyens principle).  In the CRBH geometry, the
Dirac wave function $\Psi$ behaves near infinity like solutions of the wave
equation in flat Minkowski space, and near the black hole, it should behave like
a massless particle, which decays like $t^{-\; 3/2}$ (see [12]).  Thus one would
{\it expect} (by ``interpolation") that the Dirac particle in the CRBH geometry
should decay at least as fast as $t^{-\; 3/2}$.  However, Theorem 4.2 shows that
the na\"{i}ve picture is incorrect, and that the gravitational field affects the
behavior of the Dirac particle in a far more subtle way.

Finally, one can ask why $\om = \pm \; m$ play such special roles.  To answer
this, note that for a particle of mass $m$ in flat space, its energy $\om = m$,
(or $-m$ since the Dirac equation has negative energy solutions).  If the
particle's momentum is non-zero, then $\om > m$.  Thus in flat space, its energy
is outside of the interval $(-m, m)$.  But a quantum mechanical particle can
have a continuous energy distribution, but only outside of $(-m, m)$ (this is
called the ``energy gap" or ``mass gap").  However, in the presence of a
gravitational field, the energy distribution can be supported on the entire real
line, but the dominant term is near $\pm \; m$.

We now turn to question (A), and discuss the probability for our Dirac particle
to escape to infinity, or to enter the black hole.  Let $R > \rho$, (the event
horizon), and define the probability $p$ of the Dirac particle to escape to
infinity by
\begin{equation} p = \lim_{t \ra \infty} \int_{r>R} (\bar \Psi \ga^j
\Psi)(t,x)\nu_j d\mu.  \end{equation}
Note that $p$ is independent of $R$; namely, if $R_2
> R_1 > \rho$, and \[ p_i = \lim_{t\ra \infty} \int_{r>R_i} (\bar \Psi \ga^j
\Psi)(t,x)\nu_j dx, \ \ \ i=1,2, \] then \[ p_1 - p_2 = \lim_{t\ra \infty}
\left( \int_{r>R_1} - \int_{r>R_2} \right) = \lim_{t\ra \infty} \;
\int^{R_2}_{R_1} = 0, \] by Theorem 4.1; thus $p_1 = p_2$ and this proves our
assertion.  In [12] we prove that
\[
p = \frac 1\pi
\begin{array}[t]{c}
{\displaystyle\sum_{|k|\le k_0}}\\
{\scriptstyle |n|\le n_0}
\end{array}
\int_{R\backslash[-m,m]} \left( \frac 1 2 - 2|t^{k\om n}_{12}|^2\right) \Big|
\Big< \Psi^{k\om n}_2 \; | \; \Psi_0 \Big> \Big |^2 d\om.
\]
Accordingly, $1-p$ gives the probability that the Dirac particle enters the
black hole.  The following theorem gives conditions under which $p=0, \ p=1$, or
$0 < p < 1$, in terms of the initial energy distribution.

\bigskip \noindent \underline{Theorem 4.3} ([12]): {\it Consider the Cauchy
problem as in Theorem 4.2, with initial data normalized by $\Big< \Psi_0 |
\Psi_0 \Big> = 1.$

\noindent {\rm (i)} \ \ If the outgoing initial energy distribution satisfies
$\Big< \Psi^{k\om n}_2 | \Psi_0 \Big> \not= 0$ for some $\om$ with $|\om| > m$,
then $p>0$.

\noindent {\rm(ii)} \ \ If the initial energy distribution satisfies for $a=1$
or 2, $\Big< \Psi^{k\om n}_2 | \Psi_0 \Big> \not= 0$ for some $\om, \ \ |\om|
\le m$, then $p<1$.

\noindent {\rm (iii)} \ \ If the initial energy distribution is supported in the
interval $[-m, m]$, then $p=0$.

\noindent {\rm (iv)} \ \ If (4.8) holds, then $0 < p < 1$.

\noindent {\rm (v)} \ \ $p=1$ if and only if for all $k,\om$ and $n$, the
following conditions hold:}
\begin{eqnarray*}
\Big< \Psi^{k\om n}_1 | \Psi_0 \Big> &=& 0, \ \ \ \ {\rm if} \ \ \ \ |\om| \le m
\\
\Big< \Psi^{k\om n}_1 | \Psi_0 \Big> &=& -2t^{k\om n}_{12} \Big< \Psi^{k\om n}_2
| \Psi_0 \Big>, \ \ \ {\rm if} \ \ \ \om >m.
\end{eqnarray*}

Note that in case (i), the Dirac particle has a non-zero probability of escaping
to infinity; in case (ii) the particle can enter the black hole, while in case
(iii) the components of the initial energy distribution, which lead to the decay
rate $t^{-\; 5/6}$, do not have enough energy to allow the Dirac particle to
tend to infinity.  Indeed, $\Psi^{k\om n}_2$ for $|\om| > m$ is outgoing near
the event horizon, and so the Dirac particle resists the gravitational
attraction for a while, but it eventually gets turned around and is driven into
the black hole.  In case (iv), the particle has a positive probability of either
escaping to infinity, or entering the black hole.  Finally we can understand why
$p = 1$ for special choices of initial data.  To obtain such data, consider the
special situation where a Dirac particle at time $t = -\infty$ comes in from
spatial infinity.  Taking $\Psi(0,x)$ as initial data and reversing the time
direction, the solution of this Cauchy problem clearly escapes to infinity with
probability 1.

\bigskip \bigskip \newpage
\section{Shock-Waves, Cosmology, and Black Holes}
\centerline{(joint work with B. Temple)}
\setcounter{equation}{0}

\ \ \ \ \ In the previous sections we have discussed the behavior of elementary
particles in a gravitational field; that is, how General Relativity affects
matter on small scales.  We now change directions and discuss GR on large
scales, that is we shall discuss some recent astrophysical implications of GR.

We again consider the Einstein equations (2.1) where now the energy-momentum
tensor $T_{ij}$ is that of a {\it perfect fluid}; i.e., a fluid in which
dissipation effects are neglected:
\begin{equation} T_{ij} = (p + \rho)u_iu_j + pg_{ij} . 
\end{equation}
Here $\rho$ and $p$ are the density and pressure respectively, of the fluid,
and $u =(u^0, u^1, u^2, u^3)$ is its 4-velocity, where as usual, $u_i = g_{ij}
u^j$.  The Einstein equations (5.1) describe the simultaneous evolution of the
fluid and the gravitational field: matter is the source of spacetime curvature
in Einstein's theory.

Shock-waves are relevant here since the {\it Relativistic Euler Equations} are
``embedded" within the Einstein equations $G_{ij} = \sig T_{ij}$.  This is
because the Bianchi Identities in geometry, (cf.  [23]), imply that the
(covariant) divergence of the Einstein tensor $G_{ij}$, vanishes identically:
Div $G_{ij} = 0$, so Div $T_{ij} = 0$, and this latter equation is precisely the
Relativistic Euler Equations, expressing the conservation of energy and
momentum.

We pause to make the following remarks:
\begin{itemize}
\item[(A)] There is no ``Glimm's Theorem" ([13, 17]) in GR, not yet anyway. 
However an important first step in this direction was recently made by Groah and
Temple, see [14].  \item[(B)] No proof is known which demonstrates that in GR,
shock-waves form from smooth data (see [16, 17]).  \item[(C)] In GR, the initial
data cannot be arbitrarily prescribed on a non-characteristic surface, and must
satisfy additional constraints which are imposed by geometrical considerations
(Bianchi Identities).  These constraints take the form of a coupled set of 4
nonlinear {\it elliptic equations}.  There are deep unresolved issues concerning
these constraint equations.
\end{itemize}

In [18, 19], we constructed the first examples of shock-wave solutions in GR.
The shock-waves are in the fluid variables and in the metric (gravitational
field).  This was done by matching two different spherically symmetric metrics
Lipschitz continously across a (spherically symmetric) surface of discontinuity
for the fluid variables (shock-wave).  The Inner Metric is the
Friedmann-Robertson-Walker (FRW) metric of cosmology.  Its line element is given
by
\begin{equation} ds^2 = -dt^2 + R(t)^2 \left[ \frac {dr^2}{1-kr^2} + r^2 \Big( d\theta^2 +
\sin^2 \theta d\phi^2 \Big) \right] \end{equation}
where by a suitable re-scaling of $r$,
we can take $k=0, 1, or -1$.  This metric is clearly spherically symmetric, and
describes a homogeneous, isotropic spacetime (no preferred point or direction). 
It models an expanding universe.  There is a singularity $(R=0)$ in backwards
time (i.e., earlier than present time) which corresponds to the Big-Bang.  The
function $R(t)$ determines the {\it red-shift} factor for distant astronomical
objects.  (The red-shift is what astronomers see and can measure.  It is
actually the Doppler effect applied to astrophysics).  The red shift gives
information on distance, mass, and chemical composition of the objects; see [23]
for a discussion of these things.

The Outer Metric is the Tolman-Oppenheimer-Volkoff (TOV) metric,
\begin{equation} d\bar s^2
=-B(\bar r) d\bar t^2 +\left(1 - \frac {2M(\bar r)}{\bar r}\right)^{-1} d\bar
r^2 + \bar r^2 \Big( d\theta^2 + \sin^2 \theta d\phi^2 \Big) .  \end{equation}
It is both time-independent and spherically symmetric, and models a static spacetime (or
the interior of a star).  The function $M(\bar r)$ denotes the mass inside a
ball of radius $\bar r$, and is given by \[ M(\bar r) = \int^{\bar r}_0 \ 4 \pi
s^2 \bar\rho(s) ds, \] where $\bar \rho$ is the density function.  Both the FRW
and TOV metrics satisfy Einstein's equations for a perfect fluid; see [23].

For cosmology, we assume that we are given an FRW metric (which models well our
expanding Universe), and we solve the differential equations for an unknown TOV
metric which matches the given FRW metric across an {\it expanding} (outgoing)
shock wave for the fluid variables.  Our idea is to match the given $(k=0)$ FRW
metric with equation of state given by \[ p_{{\rm matter}} \ = 0, \ \ \ p_{{\rm
radiation}} \ \ = aT^4 \ , \] where $a$ is the Stefan-Boltzmann constant, and
$T$ denotes the temperature.\footnote{This equation of state is generally taken
to be the equation of state for the Universe at present time.} In our shock-wave
model for cosmology, this model is not an expansion of the {\it entire}
universe, but rather a limited expansion into an outer ambient static spacetime
modeled by the TOV metric.

In [20], we constructed an explicit solution of this problem, and we showed that
there are unique density and pressure profiles for each initial radiation
density.  In our model, the initial Big-Bang is more like a classical
fluid-dynamical shock-wave explosion.  The mathematical theory of shock-waves,
[17] implies that many solutions decay as $t \ra \infty$ to the same shock-wave. 
Thus in our shock-wave cosmology, we cannot recover all the information about
the early details of the explosion, (due to entropy increase across
shock-waves).

However, there is a problem with our shock-wave cosmology model; namely our
shock-wave is not sufficiently far enough out.  Indeed, in [20], we calculated
the present shock position to be approximately at distance (Hubble length) \[
\frac c{H_0} \approx 10^{10} \ \ {\rm light \ years} \] away, where $H(t) =
\dot{R}/R(t)$ is the Hubble ``constant", and $H_0 = H$ (present time).  But we
observe quasars and distant galaxies at distances of $10^{11} - 10^{12}$ light
years away.  For a long time we were puzzled as to why we cannot get our
shock-wave further out.  But we have recently answered this question in a most
unexpected way.  Indeed, we recently proved that if the shock wave is beyond the
critical length scale \[ \frac c{H(t)} \ \ \ \ ({\rm Hubble \ length}) \] for
the FRW metric at any time $t > 0$, then the Universe lies ``inside a black
hole", in the sense that $(1-\frac{2M}{\bar r})$ changes sign at the critical
length scale for the FRW metric.  That is, shock matching outside a black hole
can {\it only} succeed for shocks lying inside one Hubble length from the center
of the FRW metric at any fixed time.  This is a rather surprising connection
between shock-wave cosmology and black holes.

We also have proved that radiation admitted from a galaxy (or star) at the
instant when it lies a distance of exactly one Hubble length from the FRW origin
will be observed at the origin (at a later time) to be infinitely red shifted. 
Thus an observer at the FRW origin will see the shock wave ``fading out of view"
as $t \ra \infty$.

We are in the process of modifying our shock-wave cosmology so as to allow shock
waves beyond Hubble length.

\vfill\eject
\centerline{REFERENCES}

\begin{enumerate}
\item Adler, R., Bazin, M., and Schiffer, M., ``Introduction to General
Relativity", 2nd ed.  New York: McGraw-Hill, (1975).  \item Bjorken, J., and
Drell, S., Relativistic Quantum Mechinacs, McGraw-Hill, (1964).  \item
Chandrasekhar, S., The Mathematical Theory of Black Holes, Oxford Univ.  Press,
New York, (1992).  \item Finster, F., ``Local $U(2,2)$ symmetry in relativistic
quantum mechanics", {\it J. Math.  Phys.}, {\bf 39}, 6276-6290, (1988).  \item
Finster, F., Smoller, J., Yau.  S.-T., ``Particle-like solutions of the
Einstein-Dirac equations", {\it Phys.  Rev.  D} {\bf 59}, 104020/1 -- 104020/19
(1999).  \item Finister, F., Smoller, J. , and Yau, S.-T. ``Non-existence of
black hole solutions for a spherically symmetric, static Einstein-Dirac-Maxwell
system, {\it Comm.  Math.  Phys}, {\bf 205}, 249-262, (1999).  \item Finister,
F., Smoller, J. , and Yau, S.-T. ``Non-existence of time-periodic solutions of
the Dirac equation in a Reissner-Nordstr\"{o}m black hole background", {\it J.
Math.  Phys.}, {\bf 41}, 2173-2194 (2000).  \item Finster, F., Smoller, J., Yau
S.-T., The interaction of Dirac particles with non-abelian gauge fields and
gravitation--bound states, {\it Nuclear Physics B}, {\bf 584}, 387-414 (2000). 
\item Finster, F., Smoller, J., Yau S.-T., The interaction of Dirac particles
with non-abelian gauge fields and gravitation--black holes, {\it Michigan Math. 
J.}, {\bf 47}, 199-208, (2000).  \item Finster, F., Kamran, N., Smoller, J., Yau
S.-T., Non-existence of time periodic solutions of the Dirac equation in an
axisymmetric black hole geometry, {\it Comm.  in Pure Appl.  Math.}, {\bf 53},
902-929, (2000).  \item F. Finster, N. Kamran, J. Smoller and S.-T.Yau, The long
time dynamics of Dirac particles in the Kerr-Newman black hole geometry
(preprint).  \item F. Finster, N. Kamran, J. Smoller and S.-T.Yau, Decay rates
and probability estimates Dirac particles in the Kerr-Newman black hole
geometry, {\it Comm.  Math.  Phys.}, (to appear).  \item J. Glimm, Solutions in
the large for nonlinear hyperbolic systems of equations, {\it Comm.  Pure Appl. 
Math.}, {\bf 18}, 697-715, (1965).  \item J. Groah and B. Temple, Shock-wave
solutions of the Einstein equations with perfect fluid sources: existence and
consistency for the initial-value problem, (preprint).  \item R. Kerr,
Gravitational field of a spinning body as an example of algebraically, special
metrics, {\it Phys.  Rev.  Lett.}, {\bf 11}, 237-8, (1963).  \item P. Lax,
Development of singularities of solutions of nonlinear hyperbolic partial
differential equations, {\it J. Math.  Phys.}, {\bf 5}, 611-613, (1964).  \item
Smoller, J., Shock Waves and Reaction-Diffusion Equations, 2nd ed,
Springer-Verlag, (1994).  \item Smoller, J., and Temple, B., Shock-wave
solutions of the Einstein equations, {\it Arch.  Rat.  Mech.  Anal.}, {\bf 128},
249-297, (1994).  \item Smoller, J., and Temple, B., Astrophysical shock wave
solutions of the Einstein equations, {\it Phys.  Rev.  D}, {\bf 51}, 2733-2743 ,
(1995).  \item Smoller, J., and Temple, B., Cosmology with a Shock-Wave, {\it
Comm.  Math.  Phys.}, {\bf 210}, 275-308 (2000).  \item N. Straumann and Z.
Zhou, Instability of the Bartnik-McKinnon solution of the Einstein-Yang-Mills
equations, {\it Phys.  Lett.  B.}, {\bf 237}, 353-356, (1990).  \item Wald, R.,
On the instability of the n=1 Einstein-Yang-Mills black holes and mathematically
related systems, {\it J. Math.  Phys.}, {\bf 33}, 248-255, (1992).  \item
Weinberg, S. Gravitation and Cosmology: Principles and Applications of the
General Theory of Relativity, Wiley, (1972).
\end{enumerate}

\bigskip \bigskip \noindent Mathematics Department \\
\noindent University of Michigan \\
\noindent Ann Arbor, MI 48109

\end{document}